\documentclass{elsart41}
\usepackage{graphics}
\usepackage{graphicx}
\usepackage{amssymb}
\usepackage{bm}
\begin{document}

\begin{frontmatter}



\title{Multipole ordering in $f$-electron systems}
%

\author{Katsunori Kubo\corauthref{KK}},
\author{Takashi Hotta}

\address{Advanced Science Research Center,
  Japan Atomic Energy Research Institute,
  Tokai, Ibaraki 319-1195, Japan}  

\corauth[KK]{Corresponding author.\\
}

\begin{abstract}
In order to investigate multipole ordering in $f$-electron systems
from a microscopic viewpoint,
we study the so-called $\Gamma_8$ models on three kinds of lattices,
simple cubic (sc), bcc, and fcc, based on a $j$-$j$ coupling scheme
with $f$-electron hopping integrals through $(ff\sigma)$ bonding.
From the $\Gamma_8$ model,
we derive an effective model for each lattice structure
by using the second-order perturbation theory with respect to $(ff\sigma)$.
By further applying mean-field theory to the effective model,
we find a $\Gamma_{3g}$ antiferro-quadrupole transition for the sc lattice,
a $\Gamma_{2u}$ antiferro-octupole transition for the bcc lattice,
and a longitudinal triple-$\bm{q}$ $\Gamma_{5u}$ octupole transition
for the fcc lattice.
\end{abstract}

\end{frontmatter}



In recent decades,
various kinds of magnetic and orbital ordering have been found
in $f$-electron systems.
In particular, it has been recognized that cubic systems with
$\Gamma_8$ crystalline electric field ground states frequently exhibit
higher-order multipole ordering due to their high symmetry.
Indeed, octupole ordering has been proposed to reconcile experimental
observations
for Ce$_x$La$_{1-x}$B$_6$~\cite{Kuramoto,Kusunose,Sakakibara,Kubo:CeLaB6,Kubo:CeLaB6_2}
and NpO$_2$~\cite{Santini,Santini2,Paixao,Caciuffo}.
To understand the origin of such multipole ordering from a unified view point,
it is important to analyze
a simple microscopic model with correct $f$-electron symmetry.

In this paper, we study the so-called $\Gamma_8$ models on three kinds
of lattices, simple cubic (sc), bcc, and fcc,
based on a $j$-$j$ coupling scheme.
For the description of the model, we define annihilation operators
in the second-quantized form for $\Gamma_8$ electrons
with $\alpha$ and $\beta$ orbitals as
$f_{\bm{r} \alpha \uparrow}$
=$\sqrt{5/6} a_{\bm{r} 5/2}+\sqrt{1/6} a_{\bm{r} -3/2}$,
$f_{\bm{r} \alpha \downarrow}$
=$\sqrt{5/6} a_{\bm{r} -5/2}+\sqrt{1/6} a_{\bm{r} 3/2}$,
$f_{\bm{r} \beta \uparrow}=a_{\bm{r} 1/2}$,
and $f_{\bm{r} \beta \downarrow}=a_{\bm{r} -1/2}$,
where $a_{\bm{r} j_z}$ is the annihilation operator
for an electron with the $z$-component $j_z$
of the total angular momentum $j$=5/2 at site $\bm{r}$.


In the tight-binding approximation,
the model Hamiltonian is given by~\cite{Hotta}
\begin{eqnarray}
  \mathcal{H}
  &=&
  \sum_{\bm{r},\bm{\mu},\tau,\sigma,\tau^{\prime},\sigma^{\prime}}
  t^{\bm{\mu}}_{\tau \sigma; \tau^{\prime} \sigma^{\prime}}
  f^{\dagger}_{\bm{r} \tau \sigma}
  f_{\bm{r}+\bm{\mu} \tau^{\prime} \sigma^{\prime}}
  +U \sum_{\bm{r} \tau}
  n_{\bm{r} \tau \uparrow} n_{\bm{r} \tau \downarrow}
  \nonumber \\
  &+&U^{\prime} \sum_{\bm{r}}
  n_{\bm{r} \alpha} n_{\bm{r} \beta}
  + J \sum_{\bm{r},\sigma,\sigma^{\prime}}
  f^{\dagger}_{\bm{r} \alpha \sigma}
  f^{\dagger}_{\bm{r} \beta \sigma^{\prime}}
  f_{\bm{r} \alpha \sigma^{\prime}}
  f_{\bm{r} \beta \sigma}
  \nonumber \\
  &+& J^{\prime}\sum_{\bm{r},\tau \ne \tau^{\prime}}
  f^{\dagger}_{\bm{r} \tau \uparrow}
  f^{\dagger}_{\bm{r} \tau \downarrow}
  f_{\bm{r} \tau^{\prime} \downarrow}
  f_{\bm{r} \tau^{\prime} \uparrow},
  \label{f_local}
\end{eqnarray}
where $\bm{\mu}$ is a vector connecting nearest-neighbor sites,
$t^{\bm{\mu}}_{\tau \sigma; \tau^{\prime} \sigma^{\prime}}$
is the hopping integral of an electron with
$(\tau^{\prime}, \sigma^{\prime})$ at site $\bm{r}$+$\bm{\mu}$
to the $(\tau, \sigma)$ state at $\bm{r}$
through $(ff\sigma)$ bonding~\cite{Takegahara},
$n_{\bm{r} \tau \sigma}$
=$f^{\dagger}_{\bm{r} \tau \sigma} f_{\bm{r} \tau \sigma}$,
and $n_{\bm{r} \tau}$=$\sum_{\sigma} n_{\bm{r} \tau \sigma}$.
The coupling constants $U$, $U^{\prime}$, $J$, and $J^{\prime}$
denote the intra-orbital, inter-orbital, exchange,
and pair-hopping interactions, respectively.
Note that the form of
$t^{\bm{\mu}}_{\tau \sigma; \tau^{\prime} \sigma^{\prime}}$
characterizes the lattice structure.


By using the second-order perturbation theory with respect to $(ff\sigma)$
including only the lowest energy $\Gamma_5$ triplet
among the intermediate states,
we obtain effective multipole interactions for each lattice structure.
The detail of the effective models will be reported elsewhere,
and here we report the ordered state obtained in the mean-field theory.

\begin{figure}
  \begin{center}
    \includegraphics[width=0.45\textwidth]{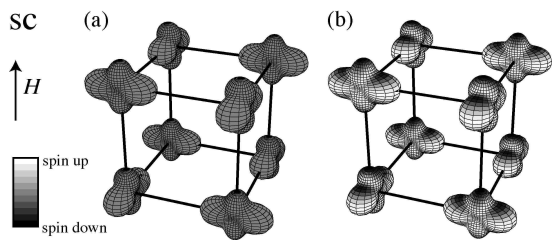}
  \end{center}
  \caption{Ordered states in the sc lattice under a magnetic field along [001].
    (a) The $\Gamma_{3g}$ AFQ state.
    (b) The FM state with the $\Gamma_{3g}$ AFQ moment.
  }
  \label{sc}
\end{figure}

For sc lattice, a $\Gamma_{3g}$ antiferro quadrupole (AFQ) transition
occurs at a finite temperature, and as lowering temperature further,
we find another transition to a ferromagnetic (FM) state.
Figures~\ref{sc}(a) and (b) show schematic views for the ordered states.
In these figures, we depict the surface defined by
$r$=$\sqrt{\sum_{\sigma}|\psi(\theta,\phi,\sigma)|^2}$
in the polar coordinates,
when the $5f$ wave-function $\Psi$ is decomposed into
radial and angular parts as
$\Psi(r, \theta, \phi, \sigma)$=$R(r)\psi(\theta,\phi,\sigma)$,
where $\sigma$ denotes real spin.
Note that white-shift in the surface indicates the increase of
the weight of up-spin state.
Figure~\ref{sc}(a) shows the AFQ state, in which
the spin distribution is isotropic
and the weights of up- and down-spin states are the same.
In the FM state state as shown in Fig.~\ref{sc}(b),
the charge distribution is the same as in Fig.~\ref{sc}(a),
while the weight of up-spin state is larger than that of
down-spin.
Note that the spin distribution is not isotropic due to
the spin-orbit interaction.

\begin{figure}
  \begin{center}
    \includegraphics[width=0.45\textwidth]{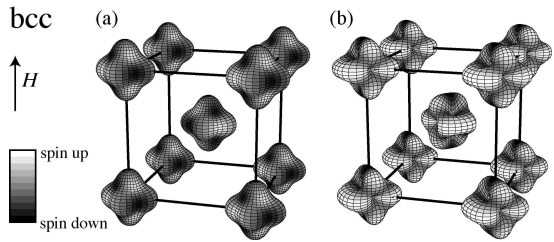}
  \end{center}
  \caption{Ordered states in the bcc lattice
    under a magnetic field along [001].
    (a) The $\Gamma_{2u}$ AFO state.
    (b) The FM state with the $\Gamma_{2u}$ AFO moment.
  }
  \label{bcc}
\end{figure}

For the bcc lattice,
a $\Gamma_{2u}$ antiferro octupole (AFO) ordering occurs first,
and then, a FM phase transition follows it.
Figure~\ref{bcc}(a) shows the $\Gamma_{2u}$ AFO state.
Since this state does not accompany any other moment,
the charge distribution retains cubic symmetry
and the weights of up- and down-spin states are the same in total.
However, the spin distribution is anisotropic,
and this state has finite octupole moment.
In the FM state as shown in Fig.~\ref{bcc}(b),
the weight of up-spin state increases,
and charge distribution is also changed
since this state has the $\Gamma_{5g}$ quadrupole moment.

\begin{figure}
  \begin{center}
    \includegraphics[width=0.28\textwidth]{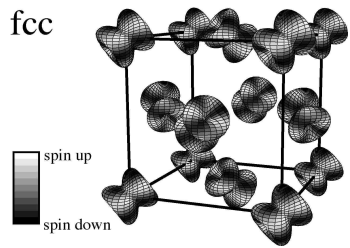}
  \end{center}
  \caption{
    The triple-$\bm{q}$ $\Gamma_{5u}$ octupole state
    in the fcc lattice.
  }
  \label{fcc}
\end{figure}

For the fcc lattice,
we have analyzed the effective model in a previous paper~\cite{Kubo:NpO2},
in which we have evaluated multipole correlation functions
by applying numerical diagonalization to the effective model
on a small cluster.
Then, we have determined that relevant interactions to the ground state
are those between $\Gamma_{4u}$ and $\Gamma_{5u}$ moments.
We have further applied mean-field theory to the simplified model
including only these relevant interactions,
and reached the conclusion that the ground state is
the longitudinal triple-$\bm{q}$ $\Gamma_{5u}$ octupole state,
which has been proposed for NpO$_2$ phenomenologically~\cite{Paixao,Caciuffo}.
Figure~\ref{fcc} shows the $\Gamma_{5u}$ octupole state.
This state accompany triple-$\bm{q}$ $\Gamma_{5g}$ quadrupole moment.
Note that this state does not have frustration even in the fcc lattice.

In summary, we have derived the multipole interaction model
from the microscopic $\Gamma_8$ Hamiltonian.
By analyzing the effective model,
we find a $\Gamma_{3g}$ AFQ state for the sc lattice,
a $\Gamma_{2u}$ AFO state for the bcc lattice,
and the longitudinal triple-$\bm{q}$ $\Gamma_{5u}$ octupole state
for the fcc lattice.


We thank S. Kambe, N. Metoki, H. Onishi, Y. Tokunaga, K. Ueda,
R. E. Walstedt, and H. Yasuoka for discussions.
K. K. is supported by the REIMEI Research Resources of
Japan Atomic Energy Research Institute.
T. H. is supported from Japan Society for the Promotion of Science
and from the Ministry of Education, Culture, Sports, Science,
and Technology of Japan

\end{document}